\documentclass[twocolumn,floatfix,showpacs]{revtex4}
\usepackage{graphicx}
\usepackage{amsmath}
\usepackage{bm}
\begin{document}

\title{Specular Andreev reflection in graphene}
\author{C. W. J. Beenakker}
\affiliation{Instituut-Lorentz, Universiteit Leiden, P.O. Box 9506, 2300 RA Leiden, The Netherlands}
\date{May 2006}
\begin{abstract}
By combining the Dirac equation of relativistic quantum mechanics with the Bogoliubov-De Gennes equation of superconductivity we investigate the electron-hole conversion at a normal-metal--superconductor interface in graphene. We find that the Andreev reflection of Dirac fermions has several unusual features: 1) The electron and hole occupy different valleys of the band structure; 2) At normal incidence the electron-hole conversion happens with unit efficiency in spite of the large mismatch in Fermi wave lengths at the two sides of the interface; and, most fundamentally: 3) Away from normal incidence the reflection angle may be the same as the angle of incidence (retro-reflection) or it may be inverted (specular reflection). Specular Andreev reflection dominates in weakly doped graphene, when the Fermi wave length in the normal region is large compared to the superconducting coherence length. We find that the transition from retro-reflection to specular reflection with decreasing doping is associated with an inversion of the voltage dependence of the subgap conductance of the interface.
\end{abstract}
\maketitle

\begin{figure}[tb]
\centerline{\includegraphics[width=0.8\linewidth]{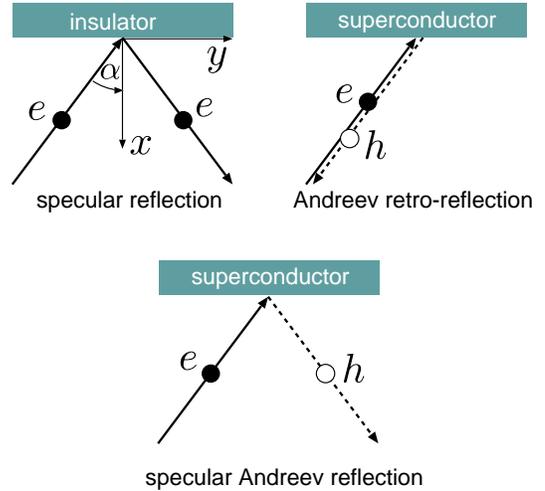}}
\caption{\label{reflections}
The top two panels show reflection processes that occur in a normal metal: specular reflection at the interface with an insulator and Andreev retro-reflection at the interface with a superconductor. Arrows indicate the direction of the velocity and solid or dashed lines distinguish whether the particle is a negatively charged electron ($e$) or a positively charged hole ($h$). In graphene there is a third possibility, indicated in the bottom panel. Like in the usual Andreev reflection, the electron is converted into a hole, but the reflection angle is inverted.
}
\end{figure}

The interface between a superconductor and a metal may reflect a negatively charged electron incident from the metal side as a positively charged hole, while the missing charge of 
$2e$ enters the superconductor as an electron pair. This electron-hole conversion, known as Andreev reflection \cite{And64}, is the process that determines the conductance of the interface at voltages below the superconducting gap --- because it is the mechanism that converts a dissipative normal current into a dissipationless supercurrent. By studying the reflection of relativistic electrons at a superconductor, we predict an unusual electron-hole conversion in graphene (a single atomic layer of carbon, with a relativistic energy spectrum \cite{Nov05,Zha05}). While usually the hole is reflected back along the path of the incident electron (retro-reflection), the Andreev reflection is specular in undoped graphene (see Fig.\ \ref{reflections}). We calculate that the difference has a clear experimental signature: The subgap conductance increases with voltage from $4/3$ to twice the ballistic value in the case of retro-reflection, but it drops from twice to $4/3$ the ballistic value in the case of specular reflection.

The practical significance of this investigation rests on the expectation that high-quality contacts between a superconductor and graphene can be realized. This expectation is supported by the experience with carbon nanotubes (rolled up sheets of graphene), which have been contacted succesfully by superconducting electrodes \cite{Kas99,Mor99,Bui03,Jar06}. The one-dimensional nature of transport in nanotubes explains why the possibility of specular Andreev reflection was not noted in that context, since it is an essentially two-dimensional effect. From a more fundamental perspective, the unusual Andreev reflection in graphene teaches us something new about the interplay of superconductivity and relativistic dynamics --- something which was not known from earlier studies of relativistic effects in heavy-element superconductors \cite{Cap95}. 

We consider a sheet of graphene in the $x-y$ plane. A superconducting electrode covers the region $x<0$ (region S), while the region $x>0$ (region N) is in the normal (non-superconducting state). Electron and hole excitations are described by the Bogoliubov-De Gennes equation \cite{Gen66},
\begin{equation}
\begin{pmatrix}
H-E_{F}&\Delta\\
\Delta^{\ast}&E_{F}-{\cal T}H{\cal T}^{-1}
\end{pmatrix}
\begin{pmatrix}
u\\ v
\end{pmatrix}=
\varepsilon
\begin{pmatrix}
u\\ v
\end{pmatrix},\label{BdGeq}
\end{equation}
with $u$ and $v$ the electron and hole wave functions, $\varepsilon>0$ the excitation energy (relative to the Fermi energy $E_{F}$), $H$ the single-particle Hamiltonian, and ${\cal T}$ the time-reversal operator. The pair potential $\Delta(\bm{r})$ couples time-reversed electron and hole states.

For $x>0$ the pair potential vanishes identically, disregarding any intrinsic superconductivity of graphene. For $x<0$ the superconducting electrode on top of the graphene layer will induce a nonzero pair potential $\Delta(x)$ via the proximity effect (similarly to what happens in a planar junction between a two-dimensional electron gas and a superconductor \cite{Vol95}). The bulk value $\Delta_{0}e^{i\phi}$ (with $\phi$ the superconducting phase) is reached at a distance from the interface which becomes negligibly small if the Fermi wave length $\lambda'_{F}$ in region S is much smaller than the value $\lambda_{F}$ in region N. We therefore adopt the step-function model
\begin{equation}
\Delta(\bm{r})=\left\{\begin{array}{ll}
\Delta_{0}e^{i\phi}&{\rm if}\;\;x<0,\\
0&{\rm if}\;\;x>0.
\end{array}\right.\label{Deltadef}
\end{equation}

We assume that the electrostatic potential $U$ in regions N and S may be adjusted independently by a gate voltage or by doping. Since the zero of potential is arbitrary, we may take
\begin{equation}
U(\bm{r})=\left\{\begin{array}{ll}
-U_{0}&{\rm if}\;\;x<0,\\
0&{\rm if}\;\;x>0.
\end{array}\right.\label{Udef}
\end{equation}
For $U_{0}$ large positive, and $E_{F}\geq 0$, the Fermi wave vector $k'_{F}\equiv 2\pi/\lambda'_{F}=(E_{F}+U_{0})/\hbar v$ in $S$ is large compared to the value $k_{F}\equiv 2\pi/\lambda_{F}=E_{F}/\hbar v$ in N (with $v$ the energy-independent velocity in graphene).

The single-particle Hamiltonian in graphene is the two-dimensional Dirac Hamiltonian \cite{Slo58},
\begin{eqnarray}
&&H=\begin{pmatrix}
H_{+}&0\\
0&H_{-}
\end{pmatrix},\label{HDiraca}\\
&&H_{\pm}=-i\hbar v(\sigma_{x}\partial_{x}\pm\sigma_{y}\partial_{y})+U,\label{HDiracb}
\end{eqnarray}
acting on a four-dimensional spinor $\bigl(\Psi_{A+},\Psi_{B+},\Psi_{A-},\Psi_{B-}\bigr)$. The indices $A,B$ label the two sublattices of the honeycomb lattice of carbon atoms, while the indices $\pm$ label the two valleys of the band structure. (There is an additional spin degree of freedom, which plays no role here.) The $2\times 2$ Pauli matrices $\sigma_{i}$ act on the sublattice index. 

The time-reversal operator interchanges the valleys \cite{Suz02},
\begin{equation}
{\cal T}=\begin{pmatrix}
0&\sigma_{z}\\
\sigma_{z}&0
\end{pmatrix}{\cal C}={\cal T}^{-1},\label{calTdef}
\end{equation}
with ${\cal C}$ the operator of complex conjugation. In the absence of a magnetic field, the Hamiltonian is time-reversal invariant, ${\cal T}H{\cal T}^{-1}=H$. Substitution into Eq.\ (\ref{BdGeq}) results in two decoupled sets of four equations each, of the form
\begin{equation}
\begin{pmatrix}
H_{\pm}-E_{F}&\Delta\\
\Delta^{\ast}&E_{F}-H_{\pm}
\end{pmatrix}
\begin{pmatrix}
u\\ v
\end{pmatrix}=
\varepsilon
\begin{pmatrix}
u\\ v
\end{pmatrix}.\label{DBdGeq}
\end{equation}
Because of the valley degeneracy it suffices to consider one of these two sets, leading to a four-dimensional Dirac-Bogoliubov-De Gennes (DBdG) equation. For definiteness we consider the set with $H_{+}$. The two-dimensional electron spinor then has components $(u_{1},u_{2})=(\Psi_{A+},\Psi_{B+})$, while the hole spinor $v={\cal T}u$ has components $(v_{1},v_{2})=(\Psi_{A-}^{\ast},-\Psi_{B-}^{\ast})$. Electron excitations in one valley are therefore coupled  by the superconductor to hole excitations in the other valley.

\begin{figure}[tb]
\centerline{\includegraphics[width=0.9\linewidth]{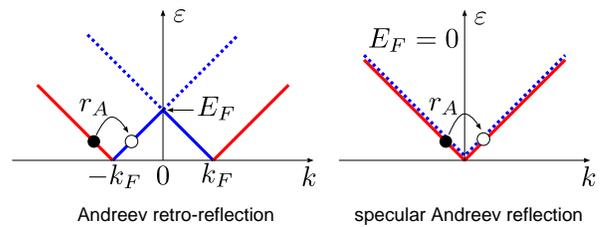}}
\caption{\label{dispersions}
Excitation spectrum in graphene, calculated from Eq.\ (\ref{dispersionlaw}) with $\Delta=0=U$ for two values of the Fermi energy $E_{F}=\hbar vk_{F}$. Red lines indicate electron excitations (filled states above the Fermi level, from one valley), while blue lines indicate hole excitations (empty states below the Fermi level, from the other valley). Solid and dotted lines distinguish the conduction and valence bands, respectively. The electron-hole conversion upon reflection at a superconductor is indicated by the arrows, for the case of normal incidence ($k\equiv k_{x}$, $k_{y}=0$). Specular Andreev reflection (right panel) happens if an electron in the conduction band is converted into a hole in the valence band. In the usual case (left panel), electron and hole both lie in the conduction band. In each case, the electron-hole conversion happens with unit probability ($|r_{A}|=1$) at normal incidence, in spite of the large wave length mismatch between the normal and superconducting regions.
}
\end{figure}

A plane wave $(u,v)\times\exp(ik_{x}x+ik_{y}y)$ is an eigenstate of the DBdG equation in a uniform system at energy
\begin{equation}
\varepsilon=\sqrt{|\Delta|^{2}+\bigl(E_{F}-U\pm \hbar v|\bm{k}|\bigr)^{2}},\label{dispersionlaw}
\end{equation}
with $|\bm{k}|=(k_{x}^{2}+k_{y}^{2})^{1/2}$. The two branches of the excitation spectrum originate from the conduction band and the valence band. The dispersion relation (\ref{dispersionlaw}) is shown in Fig.\ \ref{dispersions} for the normal region (where $\Delta=0=U$). In the superconducting region there is a gap in the spectrum of magnitude $|\Delta|=\Delta_{0}$. The mean-field requirement of superconductivity is that $\Delta_{0}\ll E_{F}+U_{0}$, or equivalently, that the superconducting coherence length $\xi=\hbar v/\Delta_{0}$ is large compared to the wave length $\lambda'_{F}$ in the superconducting region. The relative magnitude of $\xi$ and the wave length $\lambda_{F}$ in the normal region is not constrained, and we will compare the two regimes $\lambda_{F}\ll\xi_{0}$ and $\lambda_{F}\gg\xi_{0}$.

Simple inspection of the excitation spectrum shows the essential physical difference between these two regimes. Since $k_{y}$ and $\varepsilon$ are conserved upon reflection at the interface $x=0$, a general scattering state for $x>0$ is a superposition of the four $k_{x}$-values that solve Eq.\ (\ref{dispersionlaw}) at given $k_{y}$ and $\varepsilon$. The derivative $\hbar^{-1}d\varepsilon/dk_{x}$ is the expectation value $v_{x}$ of the velocity in the $x$-direction, so the reflected state contains only the two $k_{x}$-values having a positive slope. One of these two allowed $k_{x}$-values is an electron excitation ($v=0$), the other a hole excitation ($u=0$). As illustrated in Fig.\ \ref{dispersions}, the reflected hole may be either an empty state in the conduction band (for $\varepsilon<E_{F}$) or an empty state in the valence band ($\varepsilon>E_{F}$). A conduction-band hole moves opposite to its wave vector, so $v_{y}$ changes sign as well as $v_{x}$ (retro-reflection). A valence-band hole, in contrast, moves in the same direction as its wave vector, so $v_{y}$ remains unchanged and only $v_{x}$ changes sign (specular reflection). For $\varepsilon\lesssim\Delta_{0}$ the retro-reflection dominates if $E_{F}\gg \Delta_{0}$, while specular reflection dominates if $E_{F}\ll\Delta_{0}$.

To calculate the probability of the electron-hole conversion, we match a superposition of states with allowed $k_{x}$-values in N and S, demanding continuity at $x=0$. The calculation is described in the Appendix. We give the results for $\lambda'_{F}\ll \lambda_{F},\xi$, in the two regimes $E_{F}\gg\Delta_{0},\varepsilon$ and $E_{F}\ll\Delta_{0},\varepsilon$. The amplitude $r_{A}$ for Andreev reflection (from electron to hole) is
\begin{eqnarray}
&&r_{A}(\varepsilon,\alpha)=\frac{e^{-i\phi}\cos\alpha}{(\varepsilon/\Delta_{0})\cos\alpha+\zeta}, \;\;{\rm if}\;\;E_{F}\gg\varepsilon,\label{rAresulta}\\
&&r_{A}(\varepsilon,\alpha)=\frac{e^{-i\phi}\cos\alpha}{\varepsilon/\Delta_{0}+\zeta\cos\alpha},\;\;{\rm if}\;\;E_{F}\ll\varepsilon,\label{rAresultb}
\end{eqnarray}
while the amplitude $r$ for normal reflection (from electron to electron) is
\begin{eqnarray}
&&r(\varepsilon,\alpha)=\frac{-\zeta\sin\alpha}{(\varepsilon/\Delta_{0})\cos\alpha+\zeta},\;\;{\rm if}\;\;E_{F}\gg\varepsilon,\label{rresulta}\\
&&r(\varepsilon,\alpha)=\frac{-(\varepsilon/\Delta_{0})\sin\alpha}{\varepsilon/\Delta_{0}+\zeta\cos\alpha},\;\;{\rm if}\;\;E_{F}\ll\varepsilon.\label{rresultb}
\end{eqnarray}
Here $\alpha$ is the angle of incidence (as indicated in Fig.\ \ref{reflections}) and $\zeta=(\varepsilon^{2}/\Delta_{0}^{2}-1)^{1/2}$ if $\varepsilon>\Delta_{0}$, $\zeta=i(1-\varepsilon^{2}/\Delta_{0}^{2})^{1/2}$ if $\varepsilon<\Delta_{0}$. Notice that the two regimes of large and small $E_{F}$ are related by the substitution $\varepsilon/\Delta_{0}\leftrightarrow \zeta$.

One readily verifies that $|r|^{2}+|r_{A}|^{2}=1$ if $\varepsilon<\Delta_{0}$, as it should be since transmission into the superconductor is forbidden below the gap. At normal incidence ($\alpha=0$) we find $|r_{A}|^{2}=1$ for $\varepsilon<\Delta_{0}$, so the electron-hole conversion happens with unit probability. This is entirely different from usual normal-metal--superconductor junctions, where Andreev reflection is suppressed at any angle of incidence if the Fermi wave lengths at the two sides of the interface are very different. The absence of reflection without charge conversion is a consequence of the conservation of chirality (= sublattice index) by Andreev reflection: At normal incidence the incident electron and the reflected hole move on the same sublattice, while the reflection without charge conversion would require scattering from one sublattice to the other. The same conservation of chirality is responsible for the perfect transmission of normally-incident Dirac fermions through a potential barrier \cite{And98,Che06,Kat06}.

\begin{figure}[tb]
\centerline{\includegraphics[width=0.9\linewidth]{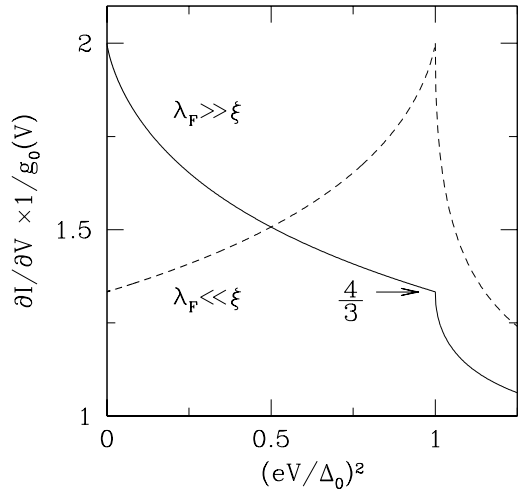}}
\caption{\label{GNS_Dirac_Point}
Differential conductance (normalized by the ballistic value $g_{0}=4Ne^2/h$) of the interface between normal and superconducting graphene, for the case of small and large Fermi wave length $\lambda_{F}$ in the normal region (relative to the coherence length $\xi=\hbar v/\Delta_{0}$ in the superconductor). The electron-hole conversion is predominantly retro-reflection for $\lambda_{F}\ll\xi$ (dashed curve), and predominantly specular reflection for $\lambda_{F}\gg\xi$ (solid curve). For $eV\leq\Delta_{0}$ the two curves are each others mirror image (when plotted versus $V^{2}$). For $eV\gg\Delta_{0}$ both curves tend to $(4-\pi)g_{0}$.
}
\end{figure}

The differential conductance of the NS junction follows from the Blonder-Tinkham-Klapwijk formula \cite{Blo82},
\begin{eqnarray}
\frac{\partial I}{\partial V}&=&g_{0}(V)\int_{0}^{\pi/2}\bigl(1-|r(eV,\alpha)|^{2}+|r_{A}(eV,\alpha)|^{2}\bigr)\nonumber\\
&&\qquad\qquad\qquad\mbox{}\times\cos\alpha\, d\alpha,\label{BTK}\\
g_{0}(V)&=&\frac{4e^{2}}{h}N(eV),\;\;N(\varepsilon)=\frac{(E_{F}+\varepsilon)W}{\pi\hbar v}.\label{g0Vdef}
\end{eqnarray}
The quantity $g_{0}$ is the ballistic conductance of $N$ transverse modes in a sheet of graphene of width $W$ (each mode having a four-fold spin and valley degeneracy). We assume $N\gg 1$, disregarding here the threshold effects that occur when $N$ becomes of order unity \cite{Kat05,Two06}. All integrals can be done analytically. The results are plotted in Fig.\ \ref{GNS_Dirac_Point}, for the two opposite regimes $\lambda_{F}\ll\xi$ and $\lambda_{F}\gg\xi$.

The differential conductance has a singularity at $eV=\Delta_{0}$, as usual for an NS junction \cite{Tin04}. For $eV\gg\Delta_{0}$ we find $\partial I/\partial V\rightarrow (4-\pi)g_{0}\approx 0.86\,g_{0}$, somewhat below the ballistic value due to the mismatch of Fermi wave lengths at the two sides of the interface. The subgap conductance, in contrast, exceeds $g_{0}$ because of Andreev reflection. The ratio $(\partial I/\partial V)/g_{0}$ varies between $4/3$ and $2$ for both retro-reflection and specular Andreev reflection, but the direction of the variation is inverted in the two cases. The difference between the solid and dashed curves in Fig.\ \ref{GNS_Dirac_Point} is a unique observable signature of the type of Andreev reflection one is dealing with.

\begin{figure}[tb]
\centerline{\includegraphics[width=0.9\linewidth]{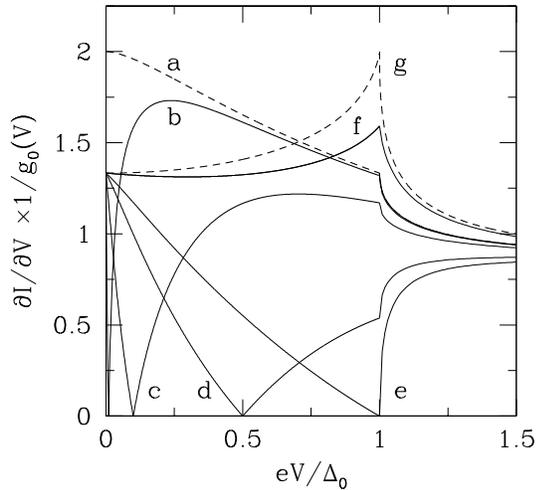}}
\caption{\label{GNS_crossover}
Differential conductance of the NS interface for $E_{F}/\Delta_{0}=0.01,0.1,0.5,1,10$ (solid curves labeled b,c,d,e,f, respectively). The dashed curves are the asymptotes for $E_{F}/\Delta_{0}\rightarrow 0,\infty$ (labeled a,g, respectively). Notice that these curves are plotted versus $V$, instead of versus $V^{2}$ as in Fig.\ \ref{GNS_Dirac_Point}.
}
\end{figure}

In experiments it may be difficult to reach the regime $E_{F}\ll\Delta_{0}$, so it is of importance to also consider the regime of comparable $E_{F}$ and $\Delta_{0}$, in which retro-reflection crosses over to specular Andreev reflection. The differential conductance in the crossover regime is plotted in Fig.\ \ref{GNS_crossover}. (See the Appendix for the calculation.) It approaches the two limiting behaviors shown in Fig.\ \ref{GNS_Dirac_Point} for $E_{F}\gg\Delta_{0}$ or $E_{F}\ll\Delta_{0}$. The crossover from one limiting curve to the other is highly nonuniform. In the limit $V\rightarrow 0$ one has $g_{0}^{-1}\partial I/\partial V\rightarrow 4/3$ for any finite ratio $E_{F}/\Delta_{0}$. For $E_{F}\leq\Delta_{0}$ the differential conductance vanishes identically at $eV=E_{F}$, because when $\varepsilon=E_{F}$ there is no Andreev reflection for any angle of incidence. These two conditions together imply a drop of $g_{0}^{-1}\partial I/\partial V$ from $4/3$ to $0$ as $eV$ increases from $0$ to $E_{F}\leq\Delta_{0}$. The drop becomes very rapid if $E_{F}\ll\Delta_{0}$. All of this should be unambiguously observable.

In conclusion, we have shown that Andreev reflection in graphene is fundamentally different from normal metals. Close to the Dirac point (at which conduction and valence bands touch), an electron from the conduction band is converted by a superconductor into a hole from the valence band. The inter-band electron-hole conversion is associated with specular reflection, instead of the usual retro-reflection (associated with electron-hole conversion within the conduction band). This is but the first example of an entirely new phenomenology to explore, regarding the interplay of superconductivity and relativistic quantum dynamics. We have demonstrated how the conductance of a single normal-superconductor interface (NS junction) is drastically changed by the transition from retro-reflection to specular Andreev reflection. We anticipate more surprises in connection with the Josephson effect for an SNS junction. 

Discussions with M. Titov are gratefully acknowledged. This research was supported by the Dutch Science Foundation NWO/FOM.

\appendix
\section{Calculation of the Andreev reflection amplitudes}

\subsection{Scattering states in N}

To calculate the Andreev reflection at the NS interface in graphene we first construct a basis of scattering states in the normal region $x>0$. These are solutions of the DBdG equation (\ref{DBdGeq}) with $\Delta=U=0$,
\begin{equation}
\begin{pmatrix}
\bm{p}\cdot\bm{\sigma}-E_{F}&0\\
0&E_{F}-\bm{p}\cdot\bm{\sigma})
\end{pmatrix}
\begin{pmatrix}
u\\ v
\end{pmatrix}=
\varepsilon
\begin{pmatrix}
u\\ v
\end{pmatrix},\label{DBdGeqN}
\end{equation}
where $\bm{p}\cdot\bm{\sigma}=-i\hbar v(\sigma_{x}\partial_{x}+\sigma_{y}\partial_{y})$.

\begin{figure}[tb]
\centerline{\includegraphics[width=0.8\linewidth]{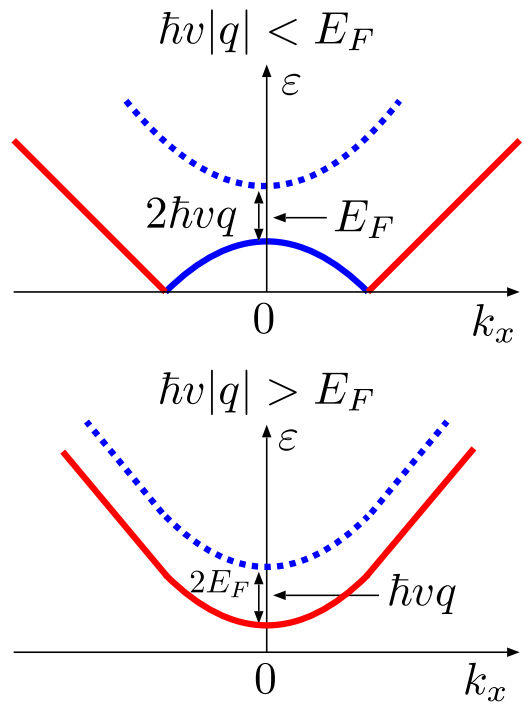}}
\caption{\label{dispersions_A}
Dispersion relation in the normal state, for two choices of transverse wave vector: $\hbar v|q|<E_{F}$ (upper panel) and $\hbar v|q|>E_{F}$ (lower panel). Electron states $\Psi^{e\pm}$ lie on the red curve, while hole states $\Psi^{h\pm}$ lie on the blue curve (solid for conduction band, dotted for valence band). The index $\pm$ is the sign of the slope, which equals the sign of the velocity in the $x$-direction.
}
\end{figure}

At a given energy $\varepsilon$ and transverse wave vector $q$ we have up to four basis states (see Fig.\ \ref{dispersions_A}),
\begin{eqnarray}
&&\Psi^{e+}=\frac{\exp(iqy+ikx)}{\sqrt{\cos\alpha}}
\begin{pmatrix}
\exp(-i\alpha/2)\\
\exp(i\alpha/2)\\
0\\
0
\end{pmatrix},\label{psiep}\\
&&\Psi^{e-}=\frac{\exp(iqy-ikx)}{\sqrt{\cos\alpha}}
\begin{pmatrix}
\exp(i\alpha/2)\\
-\exp(-i\alpha/2)\\
0\\
0
\end{pmatrix},\label{psiem}\\
&&\Psi^{h+}=\frac{\exp(iqy+ik'x)}{\sqrt{\cos\alpha'}}
\begin{pmatrix}
0\\
0\\
\exp(-i\alpha'/2)\\
-\exp(i\alpha'/2)
\end{pmatrix},\label{psihp}\\
&&\Psi^{h-}=\frac{\exp(iqy-ik'x)}{\sqrt{\cos\alpha'}}
\begin{pmatrix}
0\\
0\\
\exp(i\alpha'/2)\\
\exp(-i\alpha'/2)
\end{pmatrix},\label{psihm}
\end{eqnarray}
with the definitions
\begin{eqnarray}
&&\alpha=\arcsin[\hbar vq/(\varepsilon+E_{F})],\label{alphadef}\\
&&\alpha'=\arcsin[\hbar vq/(\varepsilon-E_{F})],\label{alphapdef}\\
&&k=(\hbar v)^{-1}(\varepsilon+E_{F})\cos\alpha,\label{kdef}\\
&&k'=(\hbar v)^{-1}(\varepsilon-E_{F})\cos\alpha'.\label{kpdef}
\end{eqnarray}

The angle $\alpha\in(-\pi/2,\pi/2)$ is the angle of incidence of the electron (having longitudinal wave vector $k$), and $\alpha'$ is the reflection angle of the hole (having longitudinal wave vector $k'$). The signs are defined such that for retro-reflection $\alpha',k'$ both have the opposite sign as $\alpha,k$ (this happens for $\varepsilon<E_{F}$), while for specular reflection $\alpha',k'$ and $\alpha,k$ have the same sign (this happens for $\varepsilon>E_{F}$). With this sign convention the states $\Psi^{e+},\Psi^{h+}$ move in the $+x$ direction (away from the NS interface), while $\Psi^{e-},\Psi^{h-}$ move in the $-x$ direction (towards the NS interface). The factors $1/\sqrt{\cos\alpha}$ and $1/\sqrt{\cos\alpha'}$ in Eqs.\ (\ref{psiep}--\ref{psihm}) ensure that all four states carry the same particle current.

There is no Andreev reflection beyond a critical angle of incidence $\alpha_{c}$, defined by
\begin{equation}
\alpha_{c}=\arcsin\left(\frac{|\varepsilon-E_{F}|}{\varepsilon+E_{F}}\right).
\end{equation}
For $|\alpha|>\alpha_{c}$ one should take
\begin{equation}
\alpha'={\rm sign}\,(\alpha)\left(\,\frac{\pi}{2}\,{\rm sign}\,(\varepsilon-E_{F})-i\,{\rm arcosh}\,\left|\frac{\sin\alpha}{\sin\alpha_{c}}\right|\right).
\end{equation}

\subsection{Scattering states in S}

\begin{figure}[tb]
\centerline{\includegraphics[width=0.8\linewidth]{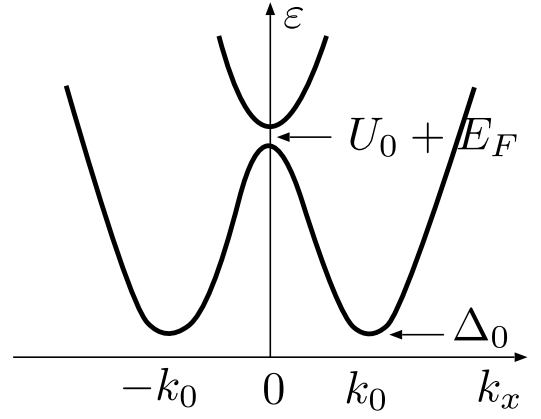}}
\caption{\label{dispersions_B}
Dispersion relation in the superconducting state, according to Eq.\ (\ref{dispersionlaw}). For $\varepsilon<\Delta_{0}$ there are no propagating waves in the superconductor.
}
\end{figure}

In the superconducting region $x<0$ we need solutions of the DBdG equation
\begin{equation}
\begin{pmatrix}
\bm{p}\cdot\bm{\sigma}-U_{0}-E_{F}&\Delta_{0}e^{i\phi}\\
\Delta_{0}e^{-i\phi}&U_{0}+E_{F}-\bm{p}\cdot\bm{\sigma})
\end{pmatrix}
\begin{pmatrix}
u\\ v
\end{pmatrix}=
\varepsilon
\begin{pmatrix}
u\\ v
\end{pmatrix}\label{DBdGeqS}
\end{equation}
that either decay as $x\rightarrow-\infty$ (for $\varepsilon<\Delta_{0}$) or which propagate in the $-x$ direction (for $\varepsilon>\Delta_{0}$). The dispersion relation (\ref{dispersionlaw}) is shown in Fig.\ \ref{dispersions_B}.

In the relevant regime $U_{0}+E_{F}\gg\Delta_{0},\varepsilon$ these solutions take the form
\begin{eqnarray}
&&\Psi^{S+}=\exp(iqy+ik_{0} x+\kappa x)\begin{pmatrix}
\exp(-i\beta)\\
\exp(i\gamma-i\beta)\\
\exp(-i\phi)\\
\exp(i\gamma-i\phi)
\end{pmatrix},\label{PsiSpdef}\\
&&\Psi^{S-}=\exp(iqy-ik_{0} x+\kappa x)\begin{pmatrix}
\exp(i\beta)\\
-\exp(-i\gamma+i\beta)\\
\exp(-i\phi)\\
-\exp(-i\gamma-i\phi)
\end{pmatrix}.\nonumber\\
&&\label{PsiSmdef}
\end{eqnarray}
The parameters $\beta,\gamma,k_{0},\kappa$ are defined by
\begin{eqnarray}
&&\beta=\left\{\begin{array}{ll}
\arccos(\varepsilon/\Delta_{0})&{\rm if}\;\;\varepsilon<\Delta_{0},\\
-i\,{\rm arcosh}\,(\varepsilon/\Delta_{0})&{\rm if}\;\;\varepsilon>\Delta_{0},
\end{array}\right. \label{betadef}\\
&&\gamma=\arcsin[\hbar vq/(U_{0}+E_{F})],\label{gammadef}\\
&&k_{0}=\sqrt{(U_{0}+E_{F})^{2}/(\hbar v)^{2}-q^{2}},\label{k0def}\\
&&\kappa=\frac{(U_{0}+E_{F})\Delta_{0}}{(\hbar v)^{2}k_{0}} \sin\beta.\label{kappadef}
\end{eqnarray}

For $\varepsilon\gg\Delta_{0}$ the state $\Psi^{S+}$ represents a hole and $\Psi^{S-}$ an electron, both propagating in the $-x$ direction. For $\varepsilon\lesssim\Delta_{0}$ these states are coherent superpositions of electron and hole excitations in the superconductor, exponentially decaying as $x\rightarrow -\infty$.

In the main text we assumed that $U_{0}\gg E_{F},\varepsilon$, meaning that the Fermi wave length $\lambda'_{F}$ in the superconducting region is much smaller than the wave length $\lambda_{F}$ in the normal region. Since $|q|\leq E_{F}/\hbar v$, this regime of a heavily doped superconductor corresponds to the limits $\gamma\rightarrow 0$, $k_{0}\rightarrow U_{0}/\hbar v$, $\kappa\rightarrow (\Delta_{0}/\hbar v)\sin\beta$. The states (\ref{PsiSpdef}) and (\ref{PsiSmdef}) thus simplify to
\begin{eqnarray}
&&\Psi^{S+}=e^{iqy+ik_{0} x+\kappa x}\begin{pmatrix}
\exp(-i\beta)\\
\exp(-i\beta)\\
\exp(-i\phi)\\
\exp(-i\phi)
\end{pmatrix},\label{PsiSpdefsimple}\\
&&\Psi^{S-}=e^{iqy-ik_{0} x+\kappa x}\begin{pmatrix}
\exp(i\beta)\\
-\exp(i\beta)\\
\exp(-i\phi)\\
-\exp(-i\phi)
\end{pmatrix}.
\label{PsiSmdefsimple}
\end{eqnarray}

\subsection{Reflection matrix}

To complete the construction of the scattering state we need to match the states in N and S at the NS boundary, by demanding continuity at $x=0$. The reflection amplitudes for an incident electron are obtained by solving
\begin{equation}
\Psi^{e-}+r\Psi^{e+}+r_{A}\Psi^{h+}=a\Psi^{S+}+b\Psi^{S-}\label{matching1}
\end{equation}
at $x=0$ for $r,r_{A},a,b$. The reflection amplitudes for an incident hole are likewise obtained by solving
\begin{equation}
\Psi^{h-}+r'\Psi^{h+}+r'_{A}\Psi^{e+}=a'\Psi^{S+}+b'\Psi^{S-}.\label{matching2}
\end{equation}
The four reflection amplitudes $r,r',r_{A},r'_{A}$ together form the reflection matrix
\begin{equation}
{\cal R}=\begin{pmatrix}
r&r'_{A}\\
r_{A}&r'
\end{pmatrix}\label{Rmatrix}
\end{equation}
of the NS interface.

We give the results in the regime $\lambda'_{F}\ll\lambda_{F}$ of a heavily doped superconductor [taking Eqs.\ (\ref{PsiSpdefsimple}-\ref{PsiSmdefsimple}) for the states in S]:
\begin{widetext}
\begin{eqnarray}
r_{A}&=&\left\{\begin{array}{ll}
e^{-i\phi}X^{-1}\sqrt{\cos\alpha\cos\alpha'}&{\rm if}\;\;|\alpha|<\alpha_{c},\\
0&{\rm if}\;\;|\alpha|>\alpha_{c},
\end{array}\right.\label{rAgeneral}\\
r&=&X^{-1}\bigl(-\cos\beta\sin[(\alpha'+\alpha)/2)]+i\sin\beta\sin[(\alpha'-\alpha)/2]\bigr),\label{rgeneral}\\
r'_{A}&=&e^{2i\phi}r_{A},\label{rApgeneral}\\
r'&=&X^{-1}\bigl(\cos\beta\sin[(\alpha'+\alpha)/2)]+i\sin\beta\sin[(\alpha'-\alpha)/2]\bigr),\label{rpgeneral}\\
X&=&\cos\beta\cos[(\alpha'-\alpha)/2)]+i\sin\beta\cos[(\alpha'+\alpha)/2].
\end{eqnarray}
\end{widetext}
One can verify the unitarity condition $({\cal R}{\cal R}^{\dagger})_{nm}=\delta_{nm}$ for $\varepsilon<\Delta_{0}$. (For $\varepsilon>\Delta_{0}$ the reflection matrix ${\cal R}$ is not unitary because there is also transmission of electrons and hole excitations into the superconductor.) One can also check that the two limits $E_{F}\gg\varepsilon$ (when $\alpha'=-\alpha$) and $E_{F}\ll\varepsilon$ (when $\alpha'=\alpha$) of Eqs.\ (\ref{rAgeneral}) and (\ref{rgeneral}) reduce to Eqs.\ (\ref{rAresulta}--\ref{rresultb}) in the main text. The full expressions given above were used to calculate the crossover curves plotted in Fig.\ \ref{GNS_crossover}.


\begin{thebibliography}{99}
\bibitem{And64} A. F. Andreev, 
Sov.\ Phys.\ JETP {\bf 19}, 1228 (1964).
\bibitem{Nov05} K. S. Novoselov, A. K. Geim, S. V. Morozov, D. Jiang,
M. I. Katsnelson, I. V. Grigorieva, S. V. Dubonos, and A. A. Firsov,
Nature {\bf 438}, 197 (2005).
\bibitem{Zha05} Y. Zhang, Y.-W. Tan, H. L. Stormer, and P. Kim,
Nature {\bf 438}, 201 (2005).
\bibitem{Kas99} A. Yu.\ Kasumov, R. Deblock, M. Kociak, B. Reulet, H. Bouchiat, I. I. Khodos, Yu.\ B. Gorbatov, V. T. Volkov, C. Journet, and M. Burghard,
Science {\bf 284}, 1508 (1999).
\bibitem{Mor99}
A. F. Morpurgo, J. Kong, C. M. Marcus, and H. Dai,
Science {\bf 286}, 263 (1999).
\bibitem{Bui03}
M. R. Buitelaar, W. Belzig, T. Nussbaumer, B. Babi\'{c}, C. Bruder, and C. Sch\"{o}nenberger,
Phys.\ Rev.\ Lett.\ {\bf 91}, 057005 (2003).
\bibitem{Jar06}
P. Jarillo-Herrero, J. A. van Dam, and L. P. Kouwenhoven,
Nature {\bf 439}, 953 (2006).
\bibitem{Cap95}
K. Capelle and E. K. U. Gross,
Phys.\ Lett.\ {\bf 198}, 261 (1995); Phys.\ Rev.\ B {\bf 59}, 7140 (1999).
\bibitem{Gen66} P. G. De Gennes, {\em Superconductivity of Metals and Alloys\/} (Benjamin, New York, 1966).
\bibitem{Vol95}
A. F. Volkov, P. H. C. Magnee, B. J. van Wees, and T. M. Klapwijk,
Physica C {\bf 242}, 261 (1995).
\bibitem{Slo58}
J. C. Slonczewski and P. R. Weiss,
Phys.\ Rev.\ {\bf 109}, 272 (1958).
\bibitem{Suz02}
H. Suzuura and T. Ando,
Phys.\ Rev.\ Lett.\ {\bf 89}, 266603 (2002).
\bibitem{And98}
T. Ando, T. Nakanishi, and R. Saito,
J. Phys.\ Soc.\ Japan, {\bf 67}, 2857 (1998).
\bibitem{Che06}
V. V. Cheianov and V. I. Fal'ko,
cond-mat/0603624.
\bibitem{Kat06}
M. I. Katsnelson, K. S. Novoselov, and A. K. Geim,
cond-mat/0604323.
\bibitem{Blo82}
G. E. Blonder, M. Tinkham, and T. M. Klapwijk,
Phys.\ Rev.\ B {\bf 25}, 4515 (1982).
\bibitem{Kat05} M. I. Katsnelson,
cond-mat/0512337.
\bibitem{Two06}
 J. Tworzyd{\l}o, B. Trauzettel, M. Titov, A. Rycerz, and C. W. J. Beenakker,
cond-mat/0603315.
\bibitem{Tin04}
M. Tinkham, {\em Introduction to Superconductivity\/} (Dover, New York, 2004).
\end{thebibliography}
\end{document}